# Sudbury Neutrino Observatory Results

A. B. McDonald
Physics Department
Queen's University
Kingston, Ontario
CANADA K7L 3N6
For the SNO Collaboration[1]

e-mail art@snolab.ca

**Abstract**

The Sudbury Neutrino Observatory uses 1000 tonnes of heavy water in an ultra-clean Cherenkov detector situated 2 km underground in Ontario, Canada to study neutrinos from the Sun and other astrophysical sources. The Charged Current (CC) reaction on deuterium is sensitive only to electron neutrinos whereas the Neutral Current (NC) is equally sensitive to all active neutrino types. By measuring the flux of neutrinos from $^8$B decay in the Sun with the CC and NC reactions it has been possible to establish clearly, through an appearance measurement, that electron neutrinos change to other active neutrino types, properties that are beyond the Standard Model of elementary particles. The observed total flux of active neutrinos agrees well with solar model flux calculations for $^8$B. This provides a clear answer to the "Solar Neutrino Problem". When these results are combined with other measurements, the oscillation of massive neutrinos is strongly defined as the primary mechanism for flavor change and oscillation parameters are well constrained.

---

[1]SNO Collaboration members are listed on references 1-5.

## 1. Introduction

The Sudbury Neutrino Observatory (SNO) [1] was designed to make separate measurements of the flux of electron neutrinos from the Sun and the flux of all active neutrino types. By a direct comparison of these two fluxes it is possible to be definitive about whether solar neutrinos are changing their types before reaching the Earth. By situating the detector 2 km underground in an active nickel mine near Sudbury, Ontario, Canada and taking extreme care with radioactivity in the construction and operation of the experiment, it has been possible to make these neutrino flux measurements with a very small contribution from background events. The results of these measurements [2,3,4,5] show clearly that about 2/3 of the electron neutrinos from $^8$B decay in the Sun change into other active neutrino types in transit to the Earth. The total $^8$B neutrino flux observed is in very good agreement with calculated [6] fluxes. Including the results of other measurements of solar [7-9] and reactor [10] neutrinos defines the oscillation of massive neutrinos as the dominant mechanism for flavor change, with mixing parameters for active neutrinos restricted to a very narrow range and oscillation to sterile neutrinos strongly disfavored. The present paper will describe results to date from the Sudbury Neutrino Observatory and indicate the physics potential for measurements planned in future.

## 2. Resolving the Solar Neutrino "Problem"

Since the pioneering measurements of Ray Davis and collaborators in the 1960's, measurements [7-9] of solar neutrinos using reactions that were solely or predominantly sensitive to electron neutrinos had observed fluxes that were factors of two or three smaller than predicted by solar model predictions [6]. This discrepancy was widely studied and came to be known as the "Solar Neutrino Problem". The experiments included radiochemical measurements with Chlorine [7] and Gallium [8] and measurements of the elastic scattering from electrons in light water Cherenkov detectors [9]. These experiments were sensitive to fluxes from different neutrino source reactions in the Sun. Solar models provided very good agreement with other properties of the Sun. However, it was not known whether the neutrino flux discrepancy was due to limitations in the solar models or unknown neutrino properties. Measurements that could distinguish the two possibilities were required.

The leading explanation arising from new neutrino physics was flavor change through the oscillation of massive neutrinos. The formalism for this process was put forward by Maki, Nakagawa, Sakata and Pontecorvo (MNSP) [11] and involved physics beyond the Standard Model of elementary particles, namely neutrino flavor change and non-zero rest mass. This theory was augmented by the possibility of enhanced oscillation through the matter-enhancement of neutrino oscillation in the Sun or Earth. [12] This process, proposed by Mikheyev and Smirnov, building on work by Wolfenstein, provided the possibility of strong enhancement, distortion of the energy spectra of oscillated neutrinos and differences in daytime and nighttime fluxes of neutrinos arising from the matter-enhancement of neutrinos passing through the Earth. These effects were sought by the experiments discussed above with no finite effects observed, even with the high sensitivity provided by the Super-Kamiokande experiment.

The Sudbury Neutrino Observatory is designed to make direct observations of solar neutrino flavor change through the following reactions in a heavy water Cherenkov detector:

$$\nu_e + d \rightarrow p + p + e^- \text{ (CC)}$$
$$\nu_x + d \rightarrow p + n + \nu_x \text{ (NC)}$$
$$\nu_x + e^- \rightarrow \nu_x + e^- \text{ (ES)}$$

The charged current (CC) reaction is sensitive only to electron neutrinos, while the neutral current (NC) reaction is sensitive to all active neutrino flavors (x = e, μ, τ) above the energy threshold of 2.2 MeV. The elastic scattering (ES) reaction is sensitive to all flavors as well, but with reduced sensitivity to $\nu_\mu$ and $\nu_\tau$. The Cherenkov light produced by the electrons in the final state is used to observe the CC and ES reactions. The NC reaction is observed through the detection of the neutron in the final state of the reaction.

By comparing the flux observed with the CC reaction (electron neutrinos) with the flux observed with the NC reaction (all active neutrino types), it is possible to make a direct measurement of neutrino flavor change. A less sensitive measurement can be made through a comparison of the CC reaction with the ES reaction. Because all reactions are sensitive to the flux of neutrinos from $^8$B decay in the Sun, the results are not sensitive to details of solar models. In fact, the flux of all neutrino types tests solar model calculations for $^8$B neutrino flux, independent of neutrino flavor change to active neutrinos. The SNO experiment has provided direct evidence of neutrino flavor change via these flux comparisons, and through the NC reaction, provided the first "appearance" observation of oscillated neutrino flavors.

## 3. The Sudbury Neutrino Observatory

The SNO detector [1] uses 1000 tonnes of ultra pure heavy water to observe Cherenkov light from the CC and ES neutrino reactions and to observe the production of neutrons by the NC reaction through three subsidiary techniques. The detector consists of about 9500 20-cm-diameter photomultipliers supported on an 18 meter diameter geodesic sphere viewing heavy water contained in a 12-meter diameter, 5-cm thick acrylic sphere. These are suspended in ultra-pure light water filling a 22-meter diameter barrel-shaped cavity coated with a polyurethane liner impermeable to water and to radon from the surrounding rock. The entire laboratory was constructed under Class 3000 clean-room conditions. All workers have showered and worn lint-free clothing during construction and operation of the experiment. The detector materials were carefully chosen for low radioactivity and extensive water purification and assay systems have been used to measure very low levels of U and Th. These radioactive backgrounds are of particular interest because the 2.4 to 2.6 MeV gamma rays from the decay chains of these naturally-occurring isotopes can photo-disintegrate deuterium. Therefore these isotopes produce a neutron background for the NC reaction, as well as low energy Cherenkov backgrounds for the CC and ES reactions. With the extreme care applied to cleanliness and material selection, it was possible to restrict the backgrounds from radioactivity to less than about 15% of the neutrino signals in all cases and to measure those backgrounds with good accuracy [1, 13]. Details of instrumental background reduction, calibration, signal extraction and systematic uncertainties are contained in references 2 to 5.

The SNO experimental plan involves three phases with different techniques for the detection of neutrons from the NC reaction. In the first phase, using pure heavy water, the Cherenkov light from the 6.25 MeV gammas produced when neutrons are captured on deuterium was used to detect the neutrons. For the second phase, about 2 tons of salt was added to the heavy water. Neutron detection efficiency was enhanced through the higher capture cross section on Cl, and the fact that the 8.6 MeV cascade of gammas produces more Cherenkov light. This light also produces a more isotropic distribution of hit PMT's than the CC events. The fluxes from the CC and NC reactions are extracted on a statistical basis with no constraint on the shape of the CC energy spectrum. For the third phase, the salt was removed and an array of $^3$He-filled proportional counters has been installed to provide neutron detection independent of the PMT array.

### 3. SNO Measurements To Date

In 2001, the SNO collaboration reported a measurement [2] of the flux of electron neutrinos from $^8$B decay in the Sun of $\Phi_{CC} = 1.59 \pm 0.07(\text{stat})^{+0.12}_{-0.11}(\text{syst}) \pm 0.05(\text{theor})$. This result is significantly lower than the measurement by the SuperKamiokande experiment [9] using the elastic scattering of neutrinos from electrons $\Phi_{ES} = 2.32 \pm 0.03(\text{stat})^{+0.08}_{-0.07}(\text{syst})$. These fluxes and those following in this section are quoted in units of $10^6$ cm$^{-2}$ sec$^{-1}$. The difference in these fluxes is 3.3 $\sigma$ from zero, providing evidence for a non-electron active component in the solar neutrino flux. In 2002, more accurate measurements were reported by SNO for the pure heavy water phase [3, 4], including data from the CC, ES and NC reactions. The non-$\nu_e$ active component was found to be $\Phi_{\mu\tau} = 3.41 \pm 0.45(\text{stat})^{+0.48}_{-0.45}(\text{syst})$, 5.3 sigma greater than zero, providing strong evidence for solar $\nu_e$ flavor transformation.

For CC events, assuming an undistorted $^8$B spectrum, the night minus day rate was found to be $14.0\% \pm 6.3\%(\text{stat})^{+1.5}_{-1.4}(\text{syst})$ of the average rate. If the total flux of active neutrinos was additionally constrained to have no asymmetry, the $\nu_e$ asymmetry was found to be $7.0\% \pm 4.9\%(\text{stat})^{+1.4}_{-1.2}(\text{syst})$.

With the addition of salt it was possible to break the strong correlation between the CC and NC event types because the pattern of light on the PMT's is different for the two reactions. An event angular distribution parameter $\beta_{14}$ (defined in reference [5]) was used to separate NC and CC events on a statistical basis. The data from the salt phase [5] are shown in Figure 1, together with the best fit to the data of the NC (8.6 MeV gamma) shape, the CC and ES reactions and a small background component determined from independent measurements. The fluxes determined from this fit are:

$$\Phi_{CC} = 1.59^{+0.08}_{-0.07}(\text{statistical})^{+0.06}_{-0.08}(\text{systematic})$$
$$\Phi_{NC} = 5.21 \pm 0.27(\text{stat}) \pm 0.38(\text{syst})$$
$$\Phi_{ES} = 2.21^{+0.31}_{-0.26}(\text{stat}) \pm 0.10(\text{syst})$$

The fluxes show clearly that about two-thirds of the electron neutrinos have changed their flavor to other active neutrino types, violating a hypothesis test for no flavor change at greater than 7 $\sigma$. The observed total flux of active neutrinos (NC) is in excellent agreement with the flux of $^8$B neutrinos obtained from solar models [6]: $\Phi_{SSM} = 5.82 \pm 1.3; 5.31 \pm 0.06$. Oscillation purely to sterile neutrinos is strongly disfavored. By comparison of the active total flux of $^8$B solar neutrinos observed by the NC reaction with the $^8$B flux calculated with solar models, restrictions are placed on sub-dominant oscillations to sterile neutrinos.

The data for solar neutrinos have been analyzed by many authors in terms of the MNSP [11] mixing matrix wherein flavor eigenstates for the three active neutrino types (l = e, $\mu$, $\tau$) are related to mass eigenstates (i) via the matrix $U_{li}$:

$$|\nu_l\rangle = \Sigma U_{li}|\nu_i\rangle.$$

The MNSP matrix $U_{li}$ can be written for three active neutrino types as:

$$U_{li} = \begin{pmatrix} c_{12} & s_{12} & 0 \\ -s_{12} & c_{12} & 0 \\ 0 & 0 & 1 \end{pmatrix} \cdot \begin{pmatrix} 1 & 0 & 0 \\ 0 & c_{23} & s_{23} \\ 0 & -s_{23} & c_{23} \end{pmatrix} \cdot \begin{pmatrix} 1 & 0 & 0 \\ 0 & 1 & 0 \\ 0 & 0 & e^{-i\delta} \end{pmatrix} \cdot \begin{pmatrix} c_{13} & 0 & s_{13} \\ 0 & 1 & 0 \\ -s_{13} & 0 & c_{13} \end{pmatrix}$$

where $c_{ij} = \cos\theta_{ij}$, and $s_{ij} = \sin\theta_{ij}$

For non-degenerate mass eigenstates, and for small $\theta_{13}$, oscillations of solar neutrinos are dominated by the first sub-matrix involving $\theta_{12}$ and this approximation is often used for analyses. The second sub-matrix dominates the oscillation of atmospheric neutrinos, the third sub-matrix involves the CP violating angle $\delta$ and the fourth sub-matrix is tested by reactor and accelerator neutrino measurements. For oscillations in vacuum in the two-neutrino mixing approximation, the survival probability for solar neutrinos with energy E, that have traveled a distance L is:

$$P(\nu_e \to \nu_e) = 1 - \sin^2 2\theta_{12} \sin^2 (1.27 \frac{\Delta m_{12}^2 L}{E})$$

If the neutrinos travel in a region of high electron density, such as in regions within the Sun or Earth, the interaction of neutrinos with electrons can produce matter enhancement of the oscillation process, via the MSW effect [12]. The MSW process produces adjustments to the effective values for $\theta_{12}$ and $\Delta m_{12}^2$, depending on the electron density. The sign of $\Delta m_{12}$ can be determined if matter interactions are substantial and distortions of the energy spectrum can occur. Interactions in the Earth can produce differences in the measured fluxes at detectors for day and night time periods.

The results for an analysis for $\nu_1$, $\nu_2$ mixing [5], following the SNO salt data is shown in Figure 2 (a). The mixing is found to be non-maximal (mixing angle less than

π/4) with a confidence level corresponding to 5.5 standard deviations. The LMA region, involving matter enhancement of the oscillation via the MSW effect is the preferred solution and that provides a determination of the sign of $\Delta m_{12}$. It is found that $m_2$ is greater than $m_1$.

Following the SNO evidence for neutrino flavor change, the KAMLAND collaboration reported their first results [10] showing the oscillation of reactor neutrinos with parameters that overlap with the solar neutrino results for part of their allowed region. Figure 2 (b) shows the allowed region when the initial Kamland data is included along with the solar data. The agreement between the solar neutrino oscillation parameters and the reactor anti-neutrino oscillation parameters confirms MNSP mass oscillations as the dominant process for flavor change. The agreement provides a confirmation of CPT for neutrinos. If full CPT invariance is assumed, the agreement confirms matter enhancement for the solar neutrinos and restricts the possibility of flavor change arising from Resonant Spin Flavor Precession [14] in the Sun. Sub-dominant transitions to sterile neutrinos and other flavor-changing mechanisms are also restricted significantly [15]. Possible processes such as neutrino decay, decoherence, violation of the equivalence principle, violation of Lorentz Invariance and flavor-changing neutral currents are strongly restricted by the observations of flavor change in solar, reactor and atmospheric neutrino experiments and by limits set in other measurements.

The principal restriction on the $\theta_{12}$ axis in Figure 2 (b) comes from the solar (mainly SNO) measurements and for the $\Delta m^2$ axis, from the KamLAND results. This arises because the $^8$B solar neutrinos in the LMA region are undergoing a resonant MSW transition and emerge from the Sun in essentially a pure $\nu_2$ state. Then, for small $\theta_{13}$, the electron neutrino survival probability (CC/NC fluxes for SNO) is almost exactly $\sin^2\theta_{12}$. On the other hand, the vacuum oscillation probability for the reactor anti-neutrinos $P(\nu_e \to \nu_e)$ is very sensitive to $\Delta m^2$.

In addition to the measurements of solar neutrinos, the SNO collaboration has reported limits on the fluxes of electron anti-neutrinos from the Sun [16] and limits on "invisible" proton decays and neutron decays that are respectively 10 and 400 times lower than previous measurements [17].

## 4. Future SNO and SNOLAB Measurements

The SNO experiment has entered its third phase with an array of $^3$He-filled proportional detectors installed in the heavy water. In this phase, the NC and CC reactions will be detected independently, removing the correlations between them and providing a more accurate measure of the NC flux. As the SNO data provide an almost direct determination of the $\theta_{12}$ parameter, the new measurements will improve the accuracy for this mixing parameter. In addition, the combination of new SNO measurements with further KamLAND reactor anti-neutrino measurements will further restrict the range of $\theta_{13}$ for low values of $\Delta m_{23}^2$.

By 2007, the sensitivity of the SNO experiment will be limited primarily by systematic uncertainties and it is planned to terminate the heavy water measurements. Discussions are underway concerning the possible replacement of the heavy water with liquid scintillator for future measurements of lower energy solar neutrinos, geo-neutrinos, reactor neutrinos and possibly double beta decay. The laboratory 2km underground is being expanded to create a long term international facility known as SNOLAB, expected to be finished in 2007. This will provide the deepest available site in future for a number of additional underground measurements. Letters of Interest have been received from over 15 international experiments proposing a very interesting future research program at SNOLAB, including measurements of solar and supernova neutrinos, dark matter and double beta decay.

## 5. Acknowledgements

This research was supported by: Canada: NSERC, Industry Canada, National Research Council, Northern Ontario Heritage Fund, Inco, AECL, Ontario Power Generation, HPCVL, CFI; US: Dept. of Energy; UK: PPARC. We thank the SNO technical staff for their strong contributions.

**Figure captions**

**Figure 1.** Distribution of (a) isotropy parameter $\beta_{14}$, (b) cos $\theta_{sun}$ and (c) Recoil electron kinetic energy, for the selected events in the fiducial volume for the salt phase of SNO. The CC and ES spectra are extracted from the data using $\beta_{14}$ and cos $\theta_{sun}$ distributions in each energy bin. Also shown are the Monte Carlo predictions for CC, ES, NC + internal and external-source neutron events, all scaled to the fit results. The dashed lines represent the summed components. All distributions are for events with effective energy > 5.5 MeV and Radius > 550 cm. Differential systematics are not shown.

**Figure 2.** Global neutrino oscillation contours for two parameter mixing. The four contour lines correspond to 90, 95, 99, 99.73 % confidence limits. (a) Solar global: $D_2O$ day and night spectra, salt CC, NC, ES fluxes, Super-Kamiokande spectral and day-night data, Cl, Ga fluxes. The best-fit point is $\Delta m^2 = 6.5 \times 10^{-5}$ eV$^2$, $\tan^2 \theta_{12} = 0.40$. (b) Solar global + KamLAND. The best-fit point is $\Delta m^2 = 67.1 \times 10^{-5}$ eV$^2$, $\tan^2 \theta_{12} = 0.41$.

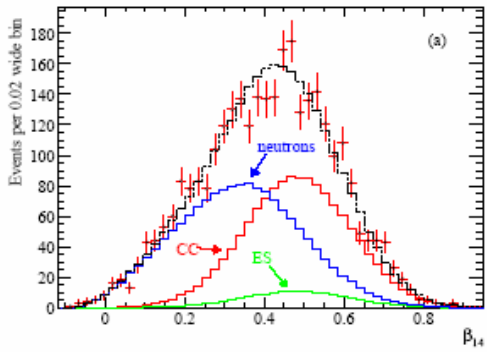
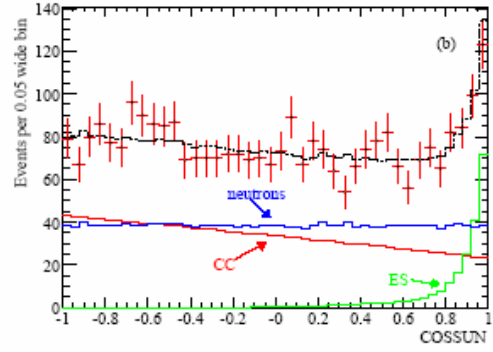
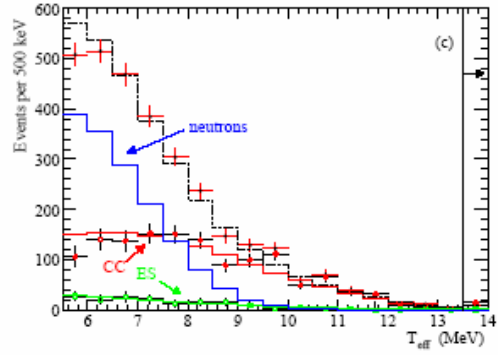
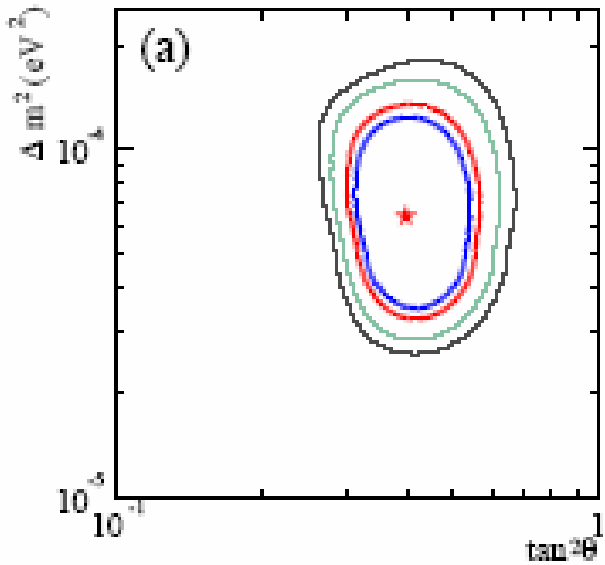
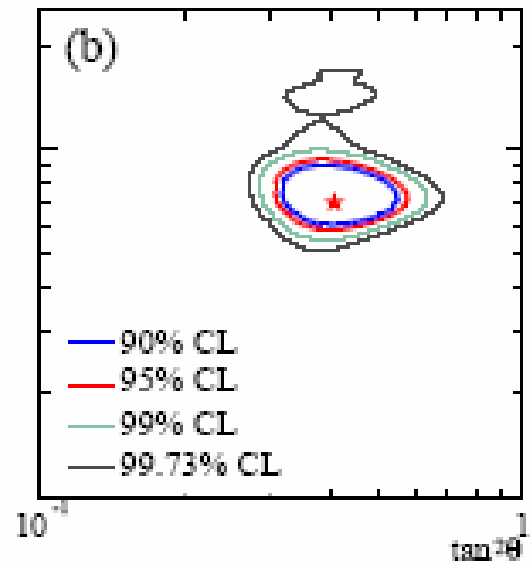